# Coupled spin states in armchair graphene nanoribbons with asymmetric zigzag edge extensions

Qiang Sun[1,†], Xuelin Yao[2,†], Oliver Gröning[1], Kristjan Eimre[1], Carlo A. Pignedoli[1], Klaus Müllen[2], Akimitsu Narita[2,3,*], Roman Fasel[1,4], and Pascal Ruffieux[1,*]

[1]*Empa, Swiss Federal Laboratories for Materials Science and Technology, 8600 Dübendorf, Switzerland*

[2]*Max Planck Institute for Polymer Research, Ackermannweg 10, 55128 Mainz, Germany*

[3]*Organic and Carbon Nanomaterials Unit, Okinawa Institute of Science and Technology Graduate University 1919-1 Tancha, Onna-son, Kunigami-gun, Okinawa 904-0495, Japan*

[4]*Department of Chemistry and Biochemistry, University of Bern, 3012 Bern, Switzerland*

*Corresponding Author: Pascal.Ruffieux@empa.ch; narita@mpip-mainz.mpg.de
† These authors contributed equally to the paper

**ABSTRACT:** Exact positioning of sublattice imbalanced nanostructures in graphene nanomaterials offers a route to control interactions between induced local magnetic moments and to obtain graphene nanomaterials with magnetically non-trivial ground states. Here, we show that such sublattice imbalanced nanostructures can be incorporated along a large band gap armchair graphene nanoribbon on the basis of asymmetric zigzag edge extensions, achieved by incorporating specifically designed precursor monomers. Scanning tunneling spectroscopy of an isolated and electronically decoupled zigzag edge extension reveals Hubbard-split states in accordance with theoretical predictions. Mean-field Hubbard-based modelling of pairs of such zigzag edge extensions reveals ferromagnetic, antiferromagnetic or quenching of the magnetic interactions depending on the relative alignment of the asymmetric edge extensions. Moreover, a ferromagnetic spin chain is demonstrated for a periodic pattern of zigzag edge extensions along the nanoribbon axis. This work opens a route towards the fabrication of graphene nanoribbon-based spin chains with complex magnetic ground states.

Interest in magnetism of graphene-based materials stems from their low spin–orbit coupling and a small natural concentration of nuclear spins of carbon, which are beneficial for long spin coherence times[1,2]. However, establishing non-trivial magnetic ground states in graphene materials requires a careful design of specific nanostructures down to atomic level and still remains a challenge[3]. According to Lieb's theorem, the ground state of a bipartite lattice possesses a total spin of $S = |N_A–N_B|/2$, where $N_A$ and $N_B$ are the number of carbon atoms belonging to sublattice A and B, respectively[4]. This simple relation provides a rule of thumb for predicting the total magnetic moments of graphene nanostructures. Recent experimental realizations of such systems include hydrogen[5] chemisorption and atomic vacancy formation[6,7] in graphene. Both structural modifications induce suppression of a $p_z$ orbital and hence an effective local sublattice imbalance, giving rise to localized spin-polarized states. Recently, small nanographenes with inherent sublattice imbalance, such as triangulene-based structures, have been synthesized, which reveal an electronic structure that is compatible with a magnetically non-trivial ground state[8–10]. For triangulene dimers as well as for Clar's goblet, a nanographene where topological frustration leads to unpaired electrons, an antiferromagnetic ground state has recently been demonstrated via inelastic spin excitation spectroscopy[11–13]. A natural next step is the incorporation of such magnetically non-trivial building blocks into structurally well-defined one-dimensional structures. Here, graphene nanoribbons (GNRs) might suit as one-dimensional scaffolds that enable atomically precise positioning of specific edge extensions hosting localized electronic states along the ribbon backbone. Interactions between these localized states could hence be precisely controlled via structural design and give rise to predictable complex magnetic ground states and excitations that solely depend on the interaction pattern of the localized states. So far, spin-polarized states are reported for zigzag-edge GNRs[14], chiral GNRs junction[15] as well as termini of armchair 7AGNRs and 5AGNR[16,17], and very recently the boron substituted GNR[18]. However, a specific structural motif inducing localized spin-polarized states within the bulk of GNRs, as is needed for the construction of GNR-based spin chains with controllably interacting magnetic moments along the GNR backbone, has still remained scarce[19–21]. Among the variety of conceivable spin chains[22,23], all with their specific scientific interest and potential technological relevance, a ferromagnetic spin chain would be particularly appealing but has remained experimentally challenging.

The realization of such systems requires the synthesis of GNR structures with atomic precision, which can be achieved via on-surface synthesis[24–27]. This bottom-up fabrication approach recently gave access to a wide range of GNRs including width-modulated GNRs with topological boundary states[19,20]. A few prototypical GNRs have furthermore been investigated on thin NaCl layers, which gives access to their intrinsic electronic properties[14,16,17,28–30]. This has been achieved through STM-based manipulation[16,31] and turns out to be particularly important for GNRs hosting localized end states where correlation-induced spin-polarization is suppressed due to the interaction with metallic growth substrates[32].

In this work, we fuse a naphtho group on the 7AGNR edge to introduce a zigzag edge extension (structure in Fig. 1a) that locally creates a sublattice imbalance of $|N_A–N_B| = 1$ along the backbone of a wide bandgap 7-atom-wide armchair GNR (7AGNR). STM-based manipulation has been applied to transfer edge-extended 7AGNRs from their Au(111) growth substrate onto NaCl islands. This allows for scanning tunneling spectroscopy (STS) characterization of their intrinsic electronic properties, which reveals spin-split states as a result of Coulomb electron-electron interactions, in qualitatively agreement with our mean-field Hubbard (MFH) modelling. Charge doping of the edge extension upon interaction with a metal surface has also been studied via STS experiments and MFH models. Importantly, various coupling configurations of the edge extensions could be realized along the 7AGNR backbone, for which different magnetic ground states are predicted depending on the relative alignment of adjacent zigzag edge extensions. We demonstrate the realization of three different configurations of zigzag edge extensions leading to either ferromagnetic, antiferromagnetic or quenching of magnetic coupling. Furthermore, fabrication of a prototypical spin chain with predicted ferromagnetic long-range order is demonstrated.

**7AGNR with asymmetric zigzag edge extension.** We first consider a 7AGNR with a single edge extension (structure highlighted by a dashed rectangle in Fig. 1a). The edge extension has an asymmetric shape with a zigzag-type head and a cove-type tail (Fig. 1a) and extends the 7AGNR backbone by 5 carbon atoms. This inevitably breaks the sublattice balance of the pristine armchair GNR backbone with $|N_A – N_B| = 1$ and is expected, according to Lieb's theorem, to induce a local magnetic moment with S = 1/2. We first examine the electronic properties of the edge-extended 7AGNR within the nearest-neighbour tight-binding (TB) model. The TB energy spectrum reveals a zero-energy state within the band gap of the 7AGNR backbone that is localized at the edge extension (Fig. 1b). In the next step, electron-electron interactions are considered by adding the on-site Coulomb repulsion U (Hubbard term, U = 3 eV) in the MFH model. As a result, the zero-energy state splits into a singly occupied state and a singly unoccupied state (with an energy gap, $\Delta_{Hub}$, of 0.34 eV) due to the high local density of states (LDOS) at the Fermi level (Fig. 1b and 1c)[33]. To fabricate such structure, we have designed a molecular precursor, 6,11-bis(10-bromoanthracen-9-yl)-1-methyltetracene (**1**), where cyclodehydrogenation and oxidative ring closure of the methyl group is expected to afford a short zigzag edge segment that smoothly bridges the anthracene and tetracene segments [19,21]. Importantly, it can be assumed that **1** can be seamlessly integrated into 7AGNRs since its conformation at the surface is expected to be closely related to 10,10'-dibromo-9,9'-bianthracene (**2**, Fig. 1d), the building block for 7AGNRs[34]. The synthesis and characterization of precursor **1** are reported in the supplementary section 1. For the on-surface synthesis of the edge-extended 7AGNRs, precursor **1** and **2** are co-deposited onto Au(111), followed by annealing to 300 °C to trigger the homolytic cleavage of the C-Br bond and induce C-C coupling, and finally oxidative cyclization (Fig. 1d)[35,36]. By adjusting the relative coverage of **1** and **2** on the surface, the density of zigzag edge extensions along the 7AGNR backbone can be controlled. A typical STM image of an isolated edge extension and a large-scale STM image are displayed in Fig. 1d and Fig. 1e, respectively. The bright protrusion in Fig. 1d is caused by the non-planar geometry of the cove edge, arising from the steric hindrance between two close-by hydrogens, and has an apparent height of ~2.6 Å (1.7 Å for the pristine 7AGNR).

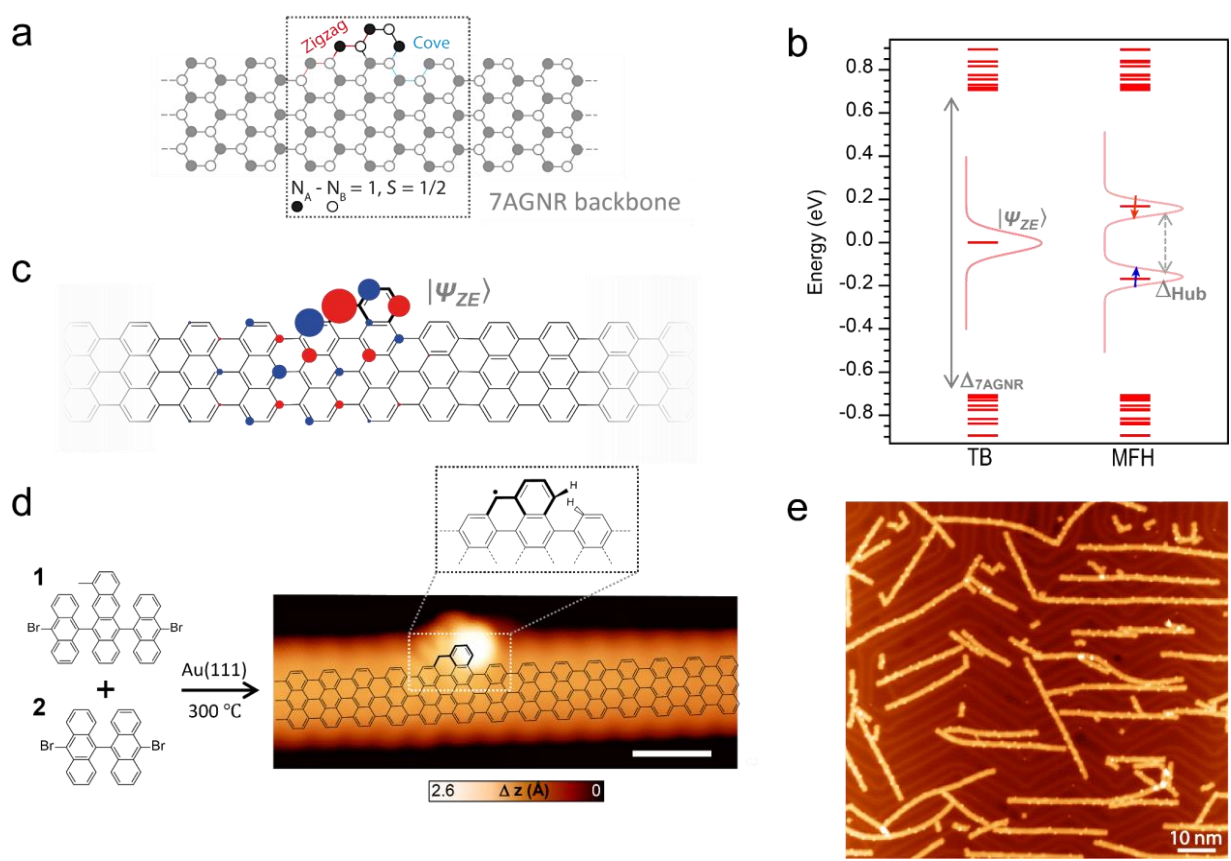

**Fig. 1 GNR edge extension and the Hubbard-split zero-energy state. a,** Schematic structure of the proposed asymmetric edge extension (indicated by a dashed rectangle) along a 7AGNR backbone,

which induces a local sublattice imbalance of 1 and a resulting total spin $S = 1/2$. Filled and open circles represent A and B sublattice atoms, respectively. The 7AGNR backbone is shown in grey. **b,** Energy spectrum of the structure shown in **a** based on nearest-neighbour TB and MFH calculations. **c,** Chemical structure of the edge extension and the wave function of the zero-energy state ($|\psi_{ZE}\rangle$). Marker size and colour indicate the wave function amplitude and parity, respectively). **d,** Synthetic route towards the edge-extended 7AGNR on Au(111), and the corresponding STM image ($V_s$ = -0.5 V, $I_t$ = 300 pA) with superimposed structural model. Scale bar: 1 nm. **e,** A large-scale STM image of a typical sample ($V_s$ = -1 V, $I_t$ = 30 pA).

To access the intrinsic electronic properties of the edge extension (i.e. to prevent charge transfer and hybridization with the electronic states of the underlying metal surface), we have initially fabricated the GNR structures on Au(111) followed by deposition of a submonolayer of NaCl. Subsequently, STM manipulation, as demonstrated earlier for pristine 7AGNRs[16], was applied to transfer a short 7AGNR exhibiting a single edge extension onto a NaCl monolayer island (Fig. 2a). Fig. 2a and 2b show STM images of the edge-extended 7AGNR before and after the manipulation-based transfer from Au(111) onto the NaCl island. A first indication for efficient electronic decoupling from the metal substrate is the increased contribution of the electronic frontier states of the edge-extended 7AGNR to the STM images. Differential conductance (dI/dV *vs* V) and current (I *vs* V) spectra have been simultaneously acquired on the edge extension to access its electronic properties. The spectra reveal a broad gap region of low conductance (current spectrum in blue) and two peaks centered at -0.2 V and 0.6 V (dI/dV spectrum in red), respectively (Fig. 2c). STM images taken at bias voltages corresponding to the positive and negative ion resonances are in good agreement with the MFH simulated LDOS maps of the Hubbard-split singly occupied and unoccupied states, respectively (Fig. 2d). We note that the comparatively low value of the Hubbard gap found by MFH model of 0.34 eV as compared to the experimental value of 0.8 eV can be due to a general underestimation of the energy gaps in MFH model and the conservative assumption of U = t = 3 eV. In the case of higher value of U = 1.6*t = 4.8 eV as suggested in a previous work[37], we find a Hubbard gap of 0.75 eV which is closer to the experimental value. For direct comparison, the calculated energy spectrum and density of states of the same edge-extended 7AGNR segment are displayed together with the experimental dI/dV spectrum in Fig. S9. It is noteworthy that the Hubbard-split states of the zigzag edge extension have the same origin as the zigzag end state of the 7AGNR[16] which stems from the local sublattice imbalance, while showing a smaller gap as compared to that of the zigzag end state of the 7AGNR on NaCl/Au(111) (0.8 eV vs 1.9 eV) (see Fig. S10). For the zigzag edge extension investigated here, the spatial extension of the spin-polarized state is estimated to be ~ 2 nm based on STM imaging (Fig. 2d), which is favorable for magnetic coupling between distant spin centers.

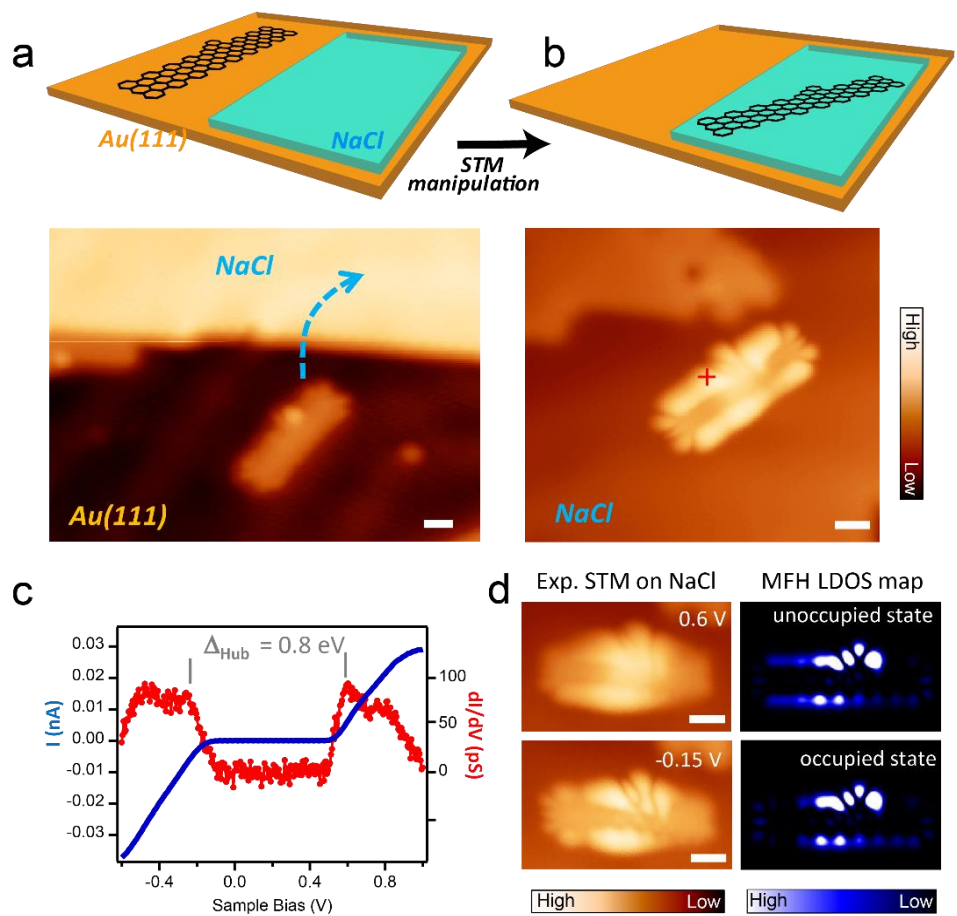

**Fig. 2 Electronic properties of the edge-extended 7AGNR on NaCl/Au(111).** Schemes and STM images **a,** before and **b,** after transferring an edge-extended 7AGNR segment from the Au(111) surface onto a NaCl insulating layer by STM manipulation (**a**: $V_s$ = -0.1 V, $I_t$ = 10 pA; **b**: $V_s$ = -0.5 V, $I_t$ = 5 pA). Scale bar: 1 nm. **c,** Differential conductance (dI/dV-V) and simultaneously acquired current (I-V) spectra recorded above the edge extension (indicated by a red cross in **b**), revealing a Hubbard gap of ~0.8 eV. **d,** STM images acquired at 0.6 V and -0.15 V, and the corresponding MFH simulated LDOS maps. Scale bar: 1 nm.

It is known that localized magnetic moments in graphene structures can be modified by chemical doping or electric fields[38,39]. We also studied the electronic properties of the edge-extended 7AGNRs on Au(111). Here, the work function difference between Au(111) (~ 5.3 eV)[40] and graphene-based materials (~ 4.5 eV)[41] entails hole doping of the zero-energy state[32,42] (a schematic drawing of the energy level alignment for GNR on Au(111) is displayed in Fig. S12). For a large enough electron/hole doping, both the spin-up and spin-down levels can be filled/emptied, and the spin-split states will thus become degenerate which leads to a nonmagnetic ground state[5,32]. The effect of hole-doping can be simulated by removing electrons in our MFH calculations. As plotted in Fig. S13a, the energy gap gradually collapses with increasing hole doping. The spin-up and spin-down levels are fully degenerate after extracting one electron from the system, resulting in a non-magnetic ground state.

This simple theoretical picture is nicely confirmed by dI/dV spectra acquired on edge-extended 7AGNRs adsorbed on Au(111) (Fig. S13b). The Hubbard-split states observed on NaCl at positive and negative bias are now degenerate, and only a single sharp peak appears just above the Fermi level (centered at 20 mV), showing that the state is unoccupied in line with recent observations[17,21]. The spectrum on the 7AGNR backbone reveals the same valence and conduction band onsets as the pristine 7AGNR[42]. The LDOS of the degenerate (positively charged) state and of the frontier bands were experimentally examined by dI/dV maps acquired at the corresponding energies (Fig. S13c). These maps agree very well with the corresponding simulated LDOS maps (Fig. S13d) of the hole-doped edge-extended 7AGNR (with one electron removed from the system).

**Coupled edge extensions and their magnetic interactions.** The spatial extension of the spin-polarized states (as estimated from Fig. 2d, i.e. 2 nm) is well suited to achieve electronic and magnetic coupling between adjacent edge extensions along the 7AGNR backbone. As fundamental building blocks in this regard, we explored the three possible dimer configurations comprising two adjacent edge extensions on opposite sides of the 7AGNR backbone, which are the ones that are exclusively observed experimentally (see Fig. S14). Figure 3a displays the three dimer configurations and indicates the corresponding overall sublattice imbalance and hence the expected local magnetic moment for each dimer. The first dimer configuration ($D_1$, cove-to-zigzag) has a sublattice imbalance of 2, which, according to Lieb's theorem, leads to S = 1, while both the second ($D_2$, cove-to-cove) and third ($D_3$, zigzag-to-zigzag) configurations have balanced sublattice occupations, yielding S = 0. The electronic structures of the different dimer configurations are investigated by both TB and MFH calculations (Fig. 3b and 3c). Two zero-energy states resulting from the edge extensions are observed in the TB energy spectrum for $D_1$ and $D_2$ configurations (with $D_2$ having a small energy splitting of ~48 meV). The on-site

Coulomb interaction included in the MFH model lifts the degeneracy of the zero-energy states and opens a gap for $D_1$ and $D_2$. Two pairs of singly occupied and unoccupied states are therefore formed with each occupied state having the same wave function as one of the unoccupied states (*cf.* a full set of wave function plots in Fig. S15), indicative for an open-shell character of dimer configurations $D_1$ and $D_2$. In clear contrast, the $D_3$ configuration does not possess zero-energy states in its TB energy spectrum. The frontier states have a substantial energy gap resulting from considerable hybridization between the zero-energy states of the involved edge extensions. This energy gap remains unchanged when considering Coulomb repulsion in the MFH approximation, corroborating a closed-shell configuration with non-equal bonding and antibonding states and absence of spin polarization (Fig. S15). For the open-shell configurations we find by MFH that $D_1$ favors ferromagnetic ordering of the magnetic moments by 12 meV, as compared to antiferromagnetic ordering (Fig. S16). This is consistent with Lieb's theorem, which predicts a triplet configuration ($S = 1$) based on the overall sublattice imbalance of $|N_A - N_B| = 2$. In clear contrast, the $D_2$ configuration favors antiferromagnetic coupling with a total spin of $S = 0$ (3 meV more stable than the ferromagnetic configuration, Fig. S16), again in line with Lieb's theorem when considering the absence of an overall sublattice imbalance for this dimer configuration. Fig. 3c and 3d illustrate the simulated wave functions of low-lying states and the spin density distributions by MFH, respectively, for all the structures considered. For the $D_3$ configuration, a strong hybridization between the two edge extensions is evidenced by the clear delocalization of the resulting wave functions, implying a closed-shell configuration. All these results are further supported by our spin unrestricted DFT calculations (Fig. S17).

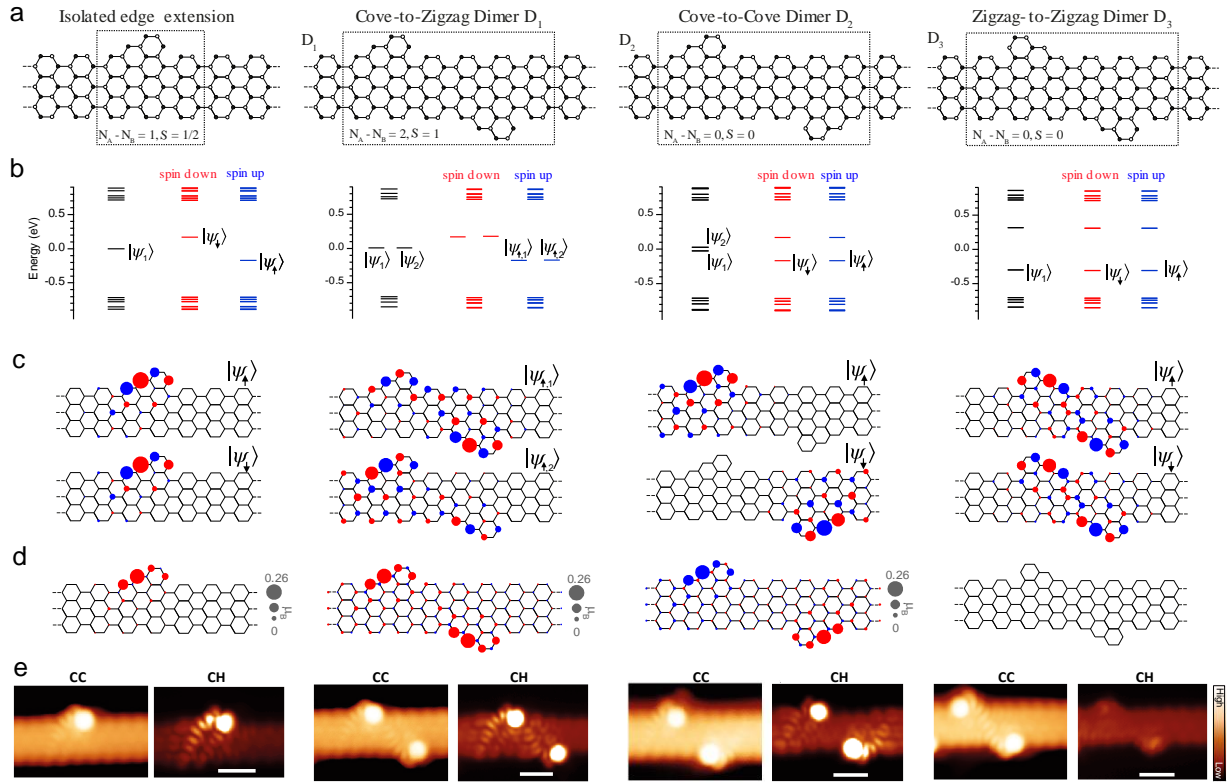

**Fig. 3 Electronic and magnetic properties of coupled zigzag edge extensions. a,** Schematic structure of an isolated edge extension and three dimer configurations (dashed boxes) embedded in a 7AGNR. Filled and open circles represent A and B sublattice atoms, respectively. *S* indicates the total spin number. **b,** Nearest-neighbour TB (black lines, U = 0 eV) and MFH energy spectrum (red and blue lines, U = 3 eV) of the corresponding structures. **c,** Wave functions of the low-energy states for the corresponding structures simulated by MFH. Red/blue markers denote opposite signs of wave functions. **d,** Spin density distribution of the corresponding structures by MFH. Red/blue isosurfaces denote spin up/spin down channels. The size of the spheres indicates the weight of spin densities (in units of the Bohr magneton $\mu_B$). **e,** Constant current (CC) STM images (from left to right: $V_s$ = -0.5 V, $I_t$ = 300 pA; $V_s$ = -1 V, $I_t$ = 500 pA; $V_s$ = -1 V, $I_t$ = 500 pA; $V_s$ = -0.85 V, $I_t$ = 550 pA) and constant height (CH) STM images ($V_t$ = 20 mV) of the corresponding structures on Au(111). Scale bars: 1 nm.

Experimentally, the three dimer structures $D_1$, $D_2$ and $D_3$ can be obtained by increasing the relative coverage of molecule **1** with respect to **2**. Fig. 3e displays STM images obtained from the corresponding structures on Au(111). Differential conductance spectra acquired over the edge extensions allow to access their electronic properties (Fig. 4a and 4b), which reveals that the electronic properties of the isolated edge extension and the $D_1$ and $D_2$ dimer configurations are very similar with one sharp peak centered around 20 mV. This observation implies that both dimers $D_1$ and $D_2$ are hole doped as in the case of the isolated extension on Au(111). On Au(111), the states that are spin-polarized in gas phase are thus degenerate due to electron transfer to the metal surface. The constant-height STM images at 20 mV display similar spatial distributions of the charge-depleted states for the single extension and the two dimer configurations $D_1$ and $D_2$ (Fig. 3e). Nonetheless, magnetic order of the two dimers should be restored upon electronic decoupling from the metal surface, exactly as in the case of the single extension (Fig. 2). On the other hand, the $D_3$ configuration reveals a clearly different spectroscopic signature with two new peaks (Fig. 4b) and absence of the state at 20 mV, as also evidenced by the constant-height STM imaging at 20 mV (Fig. 3e). This is in line with the hybridization of the two edge extension states and the formation of a hybridization gap as predicted by the calculations ($D_3$ configuration in Fig. 3). The resulting electronic structure is thus closed-shell, and non-magnetic. In Fig. S18, we examine the two hybridized states by comparing their STS maps with the simulated LDOS maps.

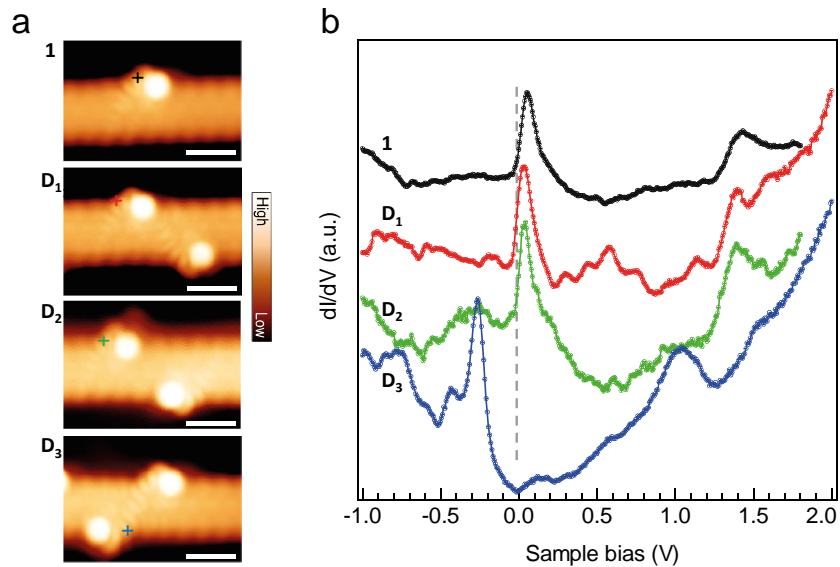

**Fig. 4 Electronic characterization of the coupled edge extensions. a,** Constant-current STM images of the isolated edge extension (1) and the three dimer configurations $D_1$, $D_2$ and $D_3$ on Au(111) (from top to bottom: $V_s$ = -0.5 V, $I_t$ = 300 pA; $V_s$ = -1 V, $I_t$ = 500 pA; $V_s$ = -1 V, $I_t$ = 500 pA; $V_s$ = -0.85 V, $I_t$ = 550 pA). Scale bars: 1 nm. **b,** dI/dV spectra taken on the four structures as indicated by markers with corresponding colours in **a**. The gray dashed line denotes the Fermi level.

Finally, we have further investigated the periodic extension of the ferromagnetically coupled $D_1$ dimer, i.e. a GNR structure with cove-to-zigzag facing edge extensions placed regularly along the 7AGNR backbone (Fig. 5a). The DFT computed electronic properties of this GNR structure in gas phase reveals an appealing ferromagnetic ground state (Fig. 5c). The GNR has a similar gap as the ferromagnetic dimer ($D_1$ configuration) but clearly dispersing frontier occupied and unoccupied states. Experimentally, such structure can be fabricated by the homocoupling of **1** followed by cyclodehydrogenation on Au(111) (Fig. 5a). A typical STM image of a sample prepared in this way is shown in Fig. 5b. Considering the synthetic approach, it is clear that the relative alignment of adjacent edge extensions cannot be controlled, but will yield a stochastic distribution of above discussed dimer configurations. Nevertheless, it is possible to find segments with periodic sequences of the $D_1$ motif and to investigate their electronic properties (STM image in Fig. 5d and dI/dV spectra in Fig. 5e). It is evident that the periodic segment exhibits one sharp state at 20 mV, in analogy to the ferromagnetically coupled dimer configuration $D_1$ discussed above. Again, hole doping due to adsorption on the Au(111) surface quenches spin polarization, which could presumably be restored with electronic decoupling of the GNR. On a rather fundamental level, long-range ferromagnetically ordered spin chains can be expected with a periodic arrangement of asymmetric zigzag edge extensions along a GNR backbone.

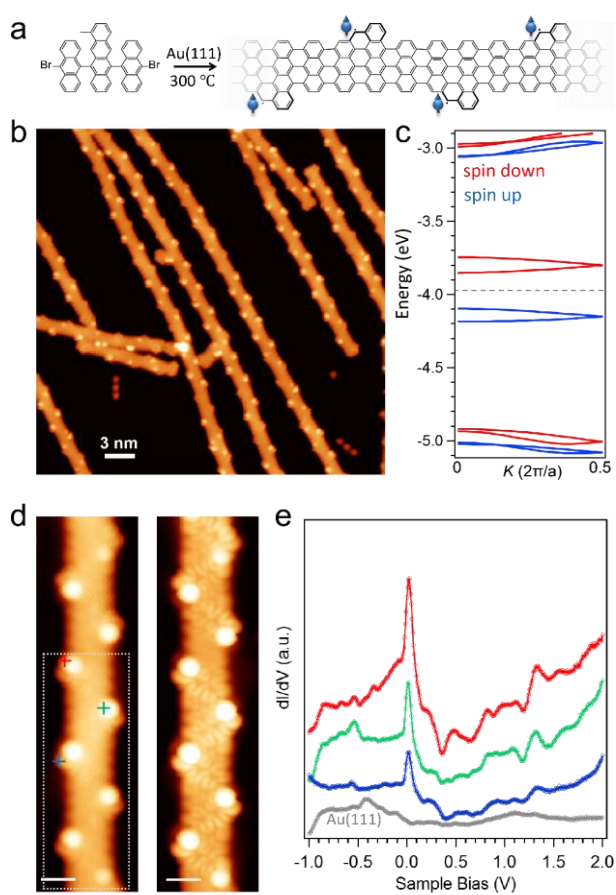

**Fig. 5 Spin chains with periodically arranged edge extensions. a,** Synthetic route towards the GNR with regularly placed edge extensions. **b,** Overview STM image of the sample ($V_s$ = -1 V, $I_t$ = 150 pA). **c,** Spin-polarized DFT calculation of the band structure of the GNR, revealing a ferromagnetic ground state. The dashed line indicates the Fermi level. **d,** Left: Close-up STM image of a GNR, with a regular segment of all $D_1$ configurations highlighted by a dashed rectangle ($V_s$ = -0.5 V, $I_t$ = 400 pA)**.** Right: Constant-current STM image acquired at 20 mV ($I_t$ = 80 pA). Scale bars: 1 nm. **e,** dI/dV spectra taken over different edge extensions on the periodic segment, as indicated by the colored crosses in **d**.

## Conclusions

In conclusion, we have synthesized 7AGNRs with fused naphtha groups, forming asymmetric zigzag edge extensions that give rise to a local sublattice imbalance of 1 and related localized zero-energy states. Electronic decoupling of a 7AGNR containing a single zigzag edge extension via STM-based manipulation onto a NaCl island reveals splitting of the zero-energy states with an energy gap of 0.8 eV. MFH and DFT simulations of the edge extension rationalize this energy splitting by correlation-induced spin-polarization with a local spin of S = 1/2. This magnetic state is quenched by hole doping upon direct adsorption on the Au(111) substrate. We have furthermore produced three different dimer configurations of coupled zigzag edge extensions which reveal ferromagnetic, antiferromagnetic or quenching of magnetic order depending on the relative orientation of the asymmetric zigzag edge extensions with respect to each other. Finally, a prototypical example of a ferromagnetic GNR spin chain comprising a regular sequence of S = 1/2 edge extensions is presented. These results are important for the understanding and further exploration of this class of new materials which may be of interest in future quantum technologies.

## Supporting information

Synthesis detail of the precursor molecules, more experimental details and methods. This material is available free of charge via the Internet at http://pubs.acs.org.

## Acknowledgements

This work was supported by the Swiss National Science Foundation under Grant No. 200020_182015 and No. 200021_172527, the NCCR MARVEL funded by the Swiss National Science Foundation (51NF40-182892), and a grant from the Swiss National Supercomputing Centre (CSCS) under project

ID s904. We are also grateful for the financial support from the Max Planck Society. Q.S. especially acknowledge the support from Yinyin Lin. X.Y. is thankful for a fellowship from the China Scholarship Council. We thank Dr. Dieter Schollmeyer for single crystal X-ray diffraction analysis.

**Author Contributions**

O.G. P.R. and R.F. conceived the experiments. Q.S. performed the on-surface synthesis and spectroscopy measurements. O.G. and Q.S. did the TB calculations. A.N., X.Y. and K.M. designed and synthesized the molecular precursors. K. E., C. A. P. and Q. S. did the DFT calculations. Q.S. and P.R. wrote the paper with contributions from all coauthors. All authors discussed the results and commented on the manuscript at all stages.

**Competing financial interests**

The authors declare no competing financial interests.

**References**

1. Droth, M. & Burkard, G. Spintronics with graphene quantum dots. *Phys. Status Solidi RRL – Rapid Res. Lett.* **10**, 75–90 (2016).

2. Balasubramanian, G. *et al.* Ultralong spin coherence time in isotopically engineered diamond. *Nat. Mater.* **8**, 383 (2009).

3. Coey, M. & Sanvito, S. The magnetism of carbon. *Phys. World* **17**, 33–37 (2004).

4. Lieb, E. H. Two theorems on the Hubbard model. *Phys. Rev. Lett.* **62**, 1201 (1989).

5. González-Herrero, H. *et al.* Atomic-scale control of graphene magnetism by using hydrogen atoms. *Science* **352**, 437–441 (2016).

6. Palacios, J. J., Fernández-Rossier, J. & Brey, L. Vacancy-induced magnetism in graphene and graphene ribbons. *Phys. Rev. B* **77**, 195428 (2008).

7. Yazyev, O. V. & Helm, L. Defect-induced magnetism in graphene. *Phys. Rev. B* **75**, 125408 (2007).

8. Pavliček, N. *et al.* Synthesis and characterization of triangulene. *Nat. Nanotechnol.* **12**, 308 (2017).

9. Su, J. *et al.* Atomically precise bottom-up synthesis of π-extended [5] triangulene. *Sci. Adv.* **5**, eaav7717 (2019).

10. Li, J. *et al.* Uncovering the Triplet Ground State of Triangular Graphene Nanoflakes Engineered with Atomic Precision on a Metal Surface. *Phys Rev Lett* **124**, 177201 (2020).

11. Yazyev, O. V. Emergence of magnetism in graphene materials and nanostructures. *Rep. Prog. Phys.* **73**, 056501 (2010).

12. Mishra, S. *et al.* Topological frustration induces unconventional magnetism in a nanographene. *Nat. Nanotechnol.* 1–7 (2019).

13. Mishra, S. *et al.* Collective All-Carbon Magnetism in Triangulene Dimers. *Angew. Chem. Int. Ed.* **59**, 12041 (2020).

14. Ruffieux, P. *et al.* On-surface synthesis of graphene nanoribbons with zigzag edge topology. *Nature* **531**, 489 (2016).

15. Li, J. *et al.* Single spin localization and manipulation in graphene open-shell nanostructures. *Nat. Commun.* **10**, 200 (2019).

16. Wang, S. *et al.* Giant edge state splitting at atomically precise graphene zigzag edges. *Nat. Commun.* **7**, 11507 (2016).

17. Lawrence, J. *et al.* Probing the Magnetism of Topological End States in 5-Armchair Graphene Nanoribbons. *ACS Nano* **14**, 4499–4508 (2020).

18. Friedrich, N. *et al.* Magnetism of topological boundary states induced by boron substitution in graphene nanoribbons. 2020, 2004.10280. ArXiv. https://arxiv.org/abs/2004.10280.

19. Gröning, O. *et al.* Engineering of robust topological quantum phases in graphene nanoribbons. *Nature* **560**, 209 (2018).

20. Rizzo, D. J. *et al.* Topological band engineering of graphene nanoribbons. *Nature* **560**, 204 (2018).

21. Rizzo, D. J. *et al.* Inducing Metallicity in Graphene Nanoribbons via Zero-Mode Superlattices. 2019, 1911.00601. ArXiv. https://arxiv.org/abs/1911.00601.

22. Affleck, I. Quantum spin chains and the Haldane gap. *J. Phys. Condens. Matter* **1**, 3047–3072 (1989).

23. Haldane, F. D. M. Excitation spectrum of a generalised Heisenberg ferromagnetic spin chain with arbitrary spin. *J. Phys. C Solid State Phys.* **15**, L1309–L1313 (1982).

24. Talirz, L., Ruffieux, P. & Fasel, R. On-Surface Synthesis of Atomically Precise Graphene Nanoribbons. *Adv. Mater.* **28**, 6222–6231 (2016).

25. Sun, Q., Zhang, R., Qiu, J., Liu, R. & Xu, W. On-Surface Synthesis of Carbon Nanostructures. *Adv. Mater.* **30**, 1705630 (2018).

26. Clair, S. & de Oteyza, D. G. Controlling a chemical coupling reaction on a surface: tools and strategies for on-surface synthesis. *Chem. Rev.* **119**, 4717–4776 (2019).

27. Li, X., Zhang, H. & Chi, L. On-Surface Synthesis of Graphyne-Based Nanostructures. *Adv. Mater.* **31**, 1804087 (2019).

28. Repp, J., Meyer, G., Stojković, S. M., Gourdon, A. & Joachim, C. Molecules on insulating films: scanning-tunneling microscopy imaging of individual molecular orbitals. *Phys. Rev. Lett.* **94**, 026803 (2005).

29. Doppagne, B. *et al.* Electrofluorochromism at the single-molecule level. *Science* **361**, 251–255 (2018).

30. Merino-Díez, N. *et al.* Unraveling the Electronic Structure of Narrow Atomically Precise Chiral Graphene Nanoribbons. *J. Phys. Chem. Lett.* **9**, 25–30 (2018).

31. Jacobse, P. H., Mangnus, M. J., Zevenhuizen, S. J. & Swart, I. Mapping the Conductance of Electronically Decoupled Graphene Nanoribbons. *ACS Nano* **12**, 7048–7056 (2018).

32. Ijäs, M. *et al.* Electronic states in finite graphene nanoribbons: Effect of charging and defects. *Phys. Rev. B* **88**, 075429 (2013).

33. Stoner, E. C. Collective electron ferronmagnetism. *Proc. R. Soc. Lond. Ser. Math. Phys. Sci.* **165**, 372–414 (1938).

34. Cai, J. *et al.* Atomically precise bottom-up fabrication of graphene nanoribbons. *Nature* **466**, 470–473 (2010).

35. Grill, L. *et al.* Nano-architectures by covalent assembly of molecular building blocks. *Nat. Nanotechnol.* **2**, 687 (2007).

36. Treier, M. *et al.* Surface-assisted cyclodehydrogenation provides a synthetic route towards easily processable and chemically tailored nanographenes. *Nat. Chem.* **3**, 61 (2011).

37. Schüler, M., Rösner, M., Wehling, T. O., Lichtenstein, A. I. & Katsnelson, M. I. Optimal Hubbard Models for Materials with Nonlocal Coulomb Interactions: Graphene, Silicene, and Benzene. *Phys Rev Lett* **111**, 036601 (2013).

38. Nair, R. *et al.* Dual origin of defect magnetism in graphene and its reversible switching by molecular doping. *Nat. Commun.* **4**, 2010 (2013).

39. Uchoa, B., Kotov, V. N., Peres, N. & Neto, A. C. Localized magnetic states in graphene. *Phys. Rev. Lett.* **101**, 026805 (2008).

40. Michaelson, H. B. The work function of the elements and its periodicity. *J. Appl. Phys.* **48**, 4729–4733 (1977).

41. Yu, Y.-J. *et al.* Tuning the graphene work function by electric field effect. *Nano Lett.* **9**, 3430–3434 (2009).

42. Ruffieux, P. *et al.* Electronic Structure of Atomically Precise Graphene Nanoribbons. *ACS Nano* **6**, 6930–6935 (2012).

# TOC Figure

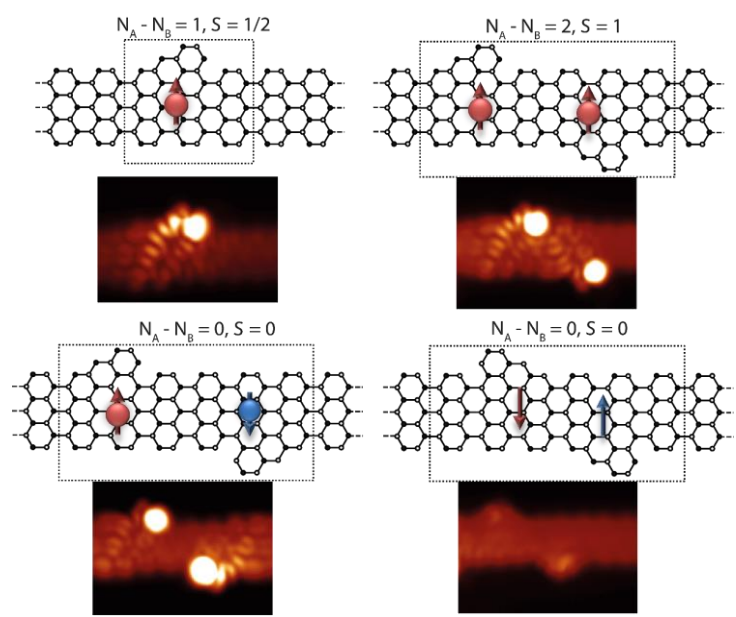